\documentclass[twocolumn,showpacs,preprintnumbers]{revtex4}

\usepackage{graphicx}
\usepackage{dcolumn}
\usepackage{bm}
\usepackage{color}
\usepackage{amsmath}
\usepackage{amsfonts}
\usepackage[pdftex]{epsfig}
\usepackage{epstopdf}

\begin{document}

\title{Study of the new superconducting compound Zr$_{0.97}$V$_{0.03}$B$_{2}$ with point-contact spectroscopy.}
\author{Rodolfo E. L\'{o}pez-Romero and R. Escudero}
\affiliation{Instituto de Investigaciones en Materiales, Universidad Nacional Aut\'{o}noma de M\'{e}xico. A. Postal 70-360. M\'{e}xico, D.F.
04510 MEXICO.}

\date{\today}

\begin{abstract}

We present a study of the superconducting energy gap performed with point contact spectroscopy in a single crystal of the superconductor Zr$_{0.97}$V$_{0.03}$B$_{2}$. We determined  the  size and evolution with temperature of the energy gap, the transition temperature in the compound occurs at the onset of T $\approx$ 9.3 K.  The  point contact junctions were measured from 1.7 to 20 K. The  differential conductance as a function of bias voltage was analyzed with the Blonder-Tinkham-Klapwijk model, the evolution of the gap with temperature was compared with the BCS theory, the ratio 2$\Delta$(0)/K$_{B}$T$_{C}$ was of the order of  8.95 indicative of a very strong coupling limit.

\end{abstract}

\pacs{74.70.Tx,74.80.Fp Superconductivity, Point Contact Spectroscopy}

\maketitle

\section{Introduction}

In new superconductors one of the important parameters to be determined  is  the transition temperature and the  emergence of the  energy gap, among other parameters. On this respect many experimental tools actually exist  to perform this determination. Perhaps  some of the  most direct tools are tunneling and point contacts spectroscopy (PCS). The first one, tunneling,  was  used in 1960 by I. Giaever to see  and measure, for the first time the superconducting energy gap in Pb \cite{IvarG, IvarGi}. Later and because the  many  experimental troubles to prepare  reliable tunnel junctions,  as frequently observed by experimentalists \cite{wolf} another spectroscopic technique quite  similar to tunneling was used;  point contact spectroscopy (PCS) developed by Sharvin \cite{Sharvin}, and used posteriorly by Yanson to study electron phonon interaction in metals and superconductors \cite{I.K.Yanson}. With this tool is  simpler  to build junctions and to  perform studies to  determinate parameters of superconductors as compared with tunneling. It is important to mention  that PCS may provide the  form to study  mechanism of quasiparticles scattered by different interactions  close to the  Fermi surface. PCS junctions are able to extend the analysis provided by  tunnel junctions  to study electronic processes in  single crystals, in different  crystallographic directions and are both complementary techniques. PCS may be used in compounds where the construction of tunnel junction is difficult. PCS and tunneling junctions analysis  mainly consist into measure  the differential resistance as a function of the bias voltage and temperature of  the material under study connected to another superconductor or to a normal electrode  \cite{I.K.Yanson, K.Yanson,Yanson}.
Many forms and  variety of techniques have  been  used  to fabricate PCS;  needle-anvil,  \cite{I.K.Yanson}, membranes with nano-holes \cite{Upadhyay}, mechanical breaking junctions \cite{Muller}.

In this work the manner  used for fabricating PCS consist of  using  a fine thin wire to  contact an edge or surface of the superconducting sample. The tip of the normal fine wire  is used to inject electrons to the superconducting sample. This procedure  was  described by Escudero et al. \cite{Escudero}. To obtain reliable spectroscopic information of the electronic parameters of the superconductor  the operating conditions for the PCS are quite  strict classified into three different regimes of conduction: ballistic, thermal, and diffuse,  established  with  the parameters associated to the material under study and  to the geometry of the  contact these are; the electronic mean free path ($\ell$) and the contact radius ($a$) \cite{Daghero,Sharvin,Duif,Jansen,Wexler}.

For the analysis of the  data measured with the PCS  various  theoretical models are used, the simplest one is the BTK model (Blonder,Tinkham and Klapwijk)\cite{Blonder}. This one describes the electronic transport  through a constriction formed by a  normal metal and a superconductor, operating  in the ballistic  regime. The  model includes the effects caused by  incorporating an insulating  barrier at the interface, the barrier allows to modeling   the limits from metallic contact to tunneling junction by varying  a parameter denoted as $Z$, the dimensionless barrier strength \cite{Blonder}. Other theoretical models based in BTK  include the degree of polarization in magnetic junctions and can be used for further analysis \cite{Strijkers}. With BTK  model is possible to determine the properties of the material under study \textsl{i.e}, the size of the energy gap, $\Delta$, its evolution with temperature, etc.
 The PCS  measurements  may give  information related to the density of electronic states,  mechanisms of electronic scattering in metals \textsl{i.e}, phonons, charge density wave, spin density, etc. Its  application to the superconducting state provides direct information on the electron-phonon interaction, responsible for the formation of the electronic coupling \cite{Michael, Terry, Jansen, Kedves, Raymond}.

In this work  we focus on  the study of the induced superconductivity in ZrB$_{2}$  by replacing Zr by  small quantities of Vanadium. The highest transition temperature  obtained at date was T$_{C} \approx$ 9.3 K,  obtained with a  substitution of 3 to 4\% of Vanadium following a procedure similar to the described by  \cite{I. K. Yanson,V. A. Gasparov,Gasparov,Soon,Renosto,J. Wang}. Our resulting composition  was  Zr$_{0.97}$V$_{0.03}$B$_{2}$.  PCS results were obtained with junctions formed with a thin  wire of tungsten plated gold of 5 $\mu$m diameter, W(Au), and a  superconducting single crystal. This study is  mainly focused into the  determination and evolution with temperature of the superconducting energy gap  analyzed using the BTK  model. At  $T=0$ K the relationship 2$\Delta/K_BT_C$ suggests a superconductor in the strong coupling limit.

\section{Experimental Details}

\subsection{Crystal Data}

Single crystals of  Zr$_{0.97}$V$_{0.03}$B$_{2}$  were obtained by mixing high purity elements and melted in a tri-arc furnace with Cu cold cathode in a inert atmosphere of Ar-H$_2$ at 5\% H$_2$. Several  stoichiometries Zr$_{1-x}$V$_x$B$_{2}$ were   formed  with values of x = 0.02, 0.04, and 0.08. Excess boron, about 10\% was introduced in order to provide the desired stoichiometry.  The resulting final composition was processed several times in order to have a homogeneous composition. At the final process  it was  annealed at high temperatures in  evacuated  quartz tubes.  The resulting were single small crystals obtained by slowly cooling the quartz container from 600$^{\circ}$C to 400$^{\circ}$C, at rates by about 1$^{\circ}$C per hour. Characteristics by X-Ray of the crystal are shown in Fig. 1 showing   the  X-ray diffraction data. The crystalline structure is type  AlB$_2$, with  space group $P6/mmm$ and  lattice parameters  determined  as  $a =3.16(5)$ \AA, and $c =3.626$ \AA,   similar to the  found by  other researchers \cite{Soon,Renosto}.

\begin{figure}[h]
	\centering
		\includegraphics[width=0.55\textwidth]{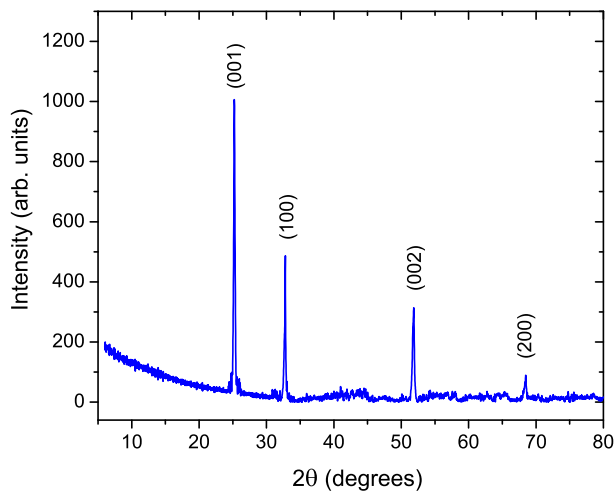}
		   \caption{(Color online) X-ray diffraction data using CuK$_{\alpha}$ radiation for the single crystal used in this work. The crystallographic parameters are $a= 3.16(5) $ \AA \textcolor[rgb]{1,1,0.88}{9}and $c =3.626$ \AA.}
	\label{Fig1.}
\end{figure}

\begin{figure}[h]
\centering
\includegraphics[width=0.5\textwidth]{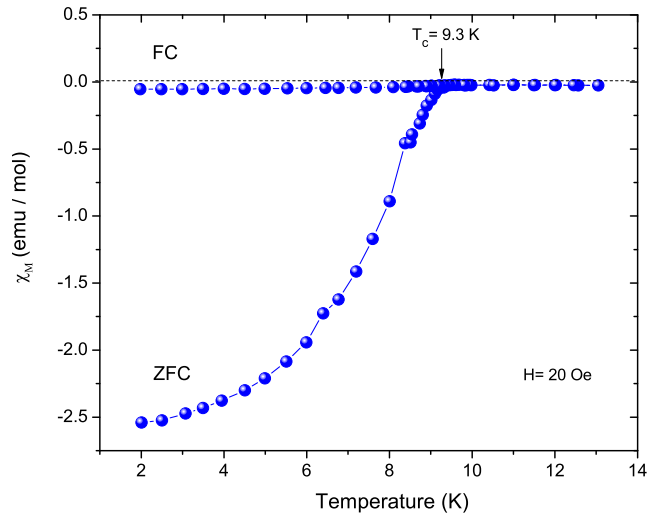}
\caption{(Color online) Molar magnetic susceptibility $\chi(T)$ for a crystal with composition  Zr$_{0.97}$V$_{0.03}$B$_{2}$ measured at 20 Oe in Zero Field Cooling and Field Cooling modes, $ZFC$ and $FC$ respectively. The amount of Meissner fraction was  close to 75\%, as determined and compared with Pb.}
\label{Fig1.}
\end{figure}

\subsection{Superconducting State}

The superconducting compound was characterized by magnetic susceptibility, $\chi (T)$, and  resistivity, $\rho (T)$, measured  in function of temperature. Magnetic susceptibility was determined in two modes Zero Field Cooling ($ZFC$) and Field Cooling ($FC$)  to determine the superconducting fraction. The  applied magnetic field was  20 Oe. The crystal used was the that present highest critical temperature onset which was about 9.3 K.  The amount of superconducting fraction determined was 75\%,  as compared with a sphere of  Pb with the same mass of the sample, see  Fig. 2.

Fig. 3 shows the resistivity as a function of temperature. The characteristic is typical of a metallic compound where the resistivity decreases with  temperature. At lower  temperature the resistivity shows  the transition to the superconducting state with  zero resistance at   $T \approx$ 9.1 K.

\begin{figure}[h]
\centering
\includegraphics[width=0.54\textwidth]{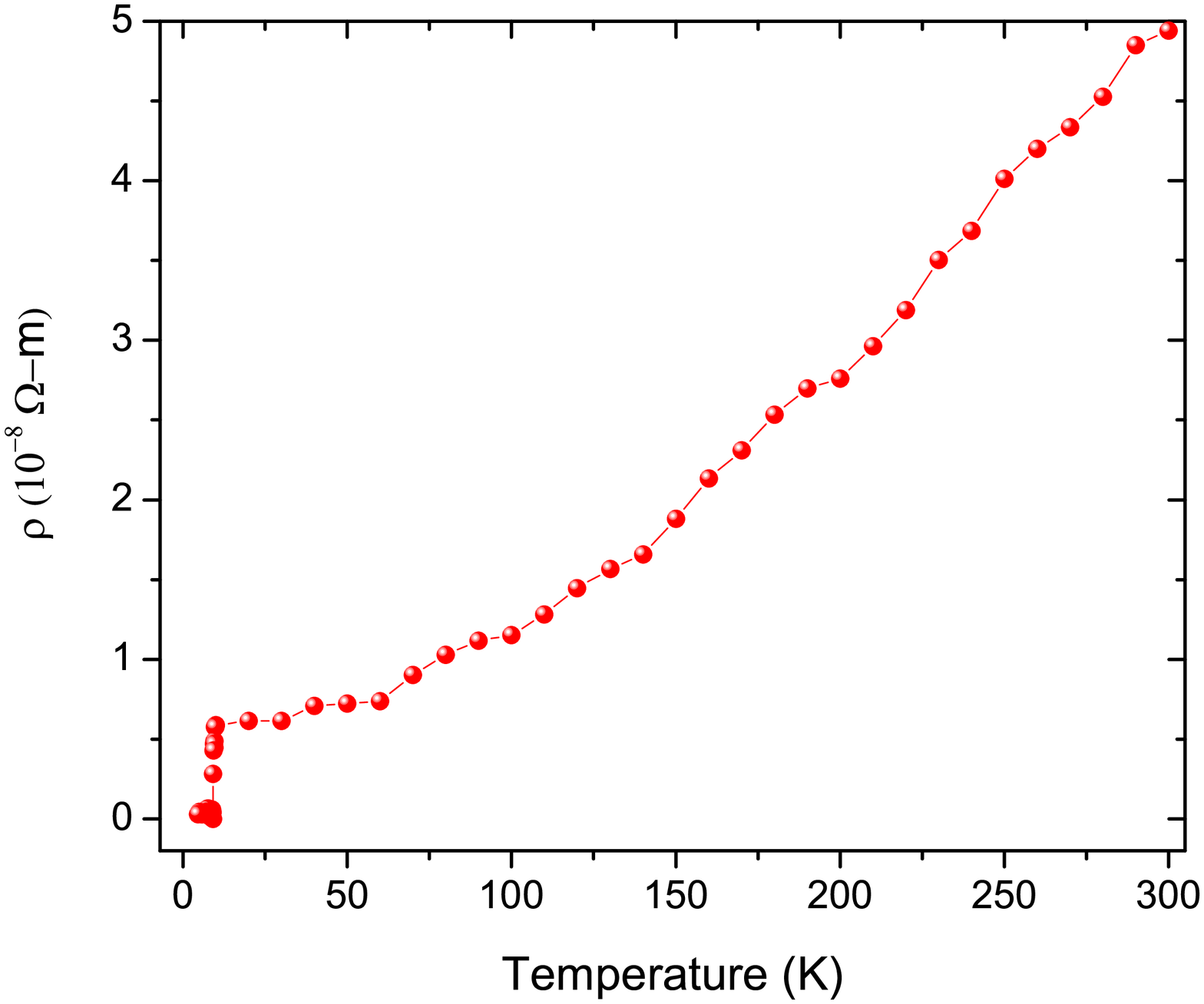}
\caption{(Color online) Resistivity as a function of temperature, $\rho -T$, of the compound Zr$_{0.97}$V$_{0.03}$B$_{2}$. The transition temperature onset is 9.3 K, and the zero resistivity is 9.1 K.}
\label{Fig1.}
\end{figure}

\subsection{Point Contact Spectroscopy}

With PCS  junctions we were able to study the superconducting state,  the contacts were prepared  in two distinct manner: crossing a W(Au) thin wire of 5 $\mu$m diameter over one  edge of the crystal,  or on a flat surface of it. The sample  was glued  to  a glass substrate with Oxford varnish in order to have good thermal contact. The  normal electrode wire (W(Au)) was  stretched over the sample to  touch  an edge  or a flat surface of the crystal.

   As mentioned before, several conditions need to be fulfilled  when measuring  PCS, one is the  working regime. Whereas smaller the area of the contact more  relevant (the spectroscopic characteristics)   will be. The most reliable electronic characteristics  will be obtain when heating effects are completely eliminated or are small \cite{Blonder, Daghero, Duif, Jansen, Branislav}.

Accordingly,  the best working conditions to study spectroscopic characteristics will be when a junction is in the ballistic regime. In it the injected electrons only will be scattered into  the studied compound by elementary excitations; phonons, magnons, etc. \cite{Jansen, Escudero, REscudero, Sinchenko}. We have to note that this regime is quite difficult to achieve,   because the size of the contact must be much smaller than  the electronic mean free path of the compound  under study $a \ll \ell$. This regime in many circumstances may be practically impossible to achieve considering that in superconducting materials frequently the mean free path ($\ell$) may be  very small of the order of  a few nanometers \cite{Daghero, Duif, Michael, Terry, Jansen, Blonder, Strijkers, Kedves, Raymond, Branislav}.

Also, we have to mention that at this moment we do not have information about the size of the electronic means free path of this compound. But however,  because the values of PCS differential resistances determined at zero bias voltage, which are of the order of 30 - 50 $\Omega$,  we may estimated that our experimental determination of the spectroscopic characteristics  will be located close to the  ballistic regime and in the worst case  into  the diffuse limit but far away to the Maxwell regime,  where heating effect are  very important and the spectroscopic features are totally  distorted. All   this consideration were taking into account according as mentioned by  Yanson, et al. \cite{yanson}. Lastly another important condition to be fulfilled and used to characterize the type of junction is related to the $Z$ parameter given by  the BTK model. It tell us if  the  spectroscopic characteristic are type  tunnel or PCS. The size of this parameter was determined in  our  studies at  $Z \approx$ 3.

The electronic equipment  used to characterize the junctions and to  obtain, first  the  differential resistance,  and after the differential conductance  as functions  of the bias voltage ($dV/dI$, and $dI/dV$ vs $V$, respectively) was the $ac$ modulation technique with lock-in amplifier and  bridge {\cite{wolf, Adler}  and a MPMS Quantum Design system as a cryostat.

\begin{figure}[h]
\centering
\includegraphics[width=0.5\textwidth]{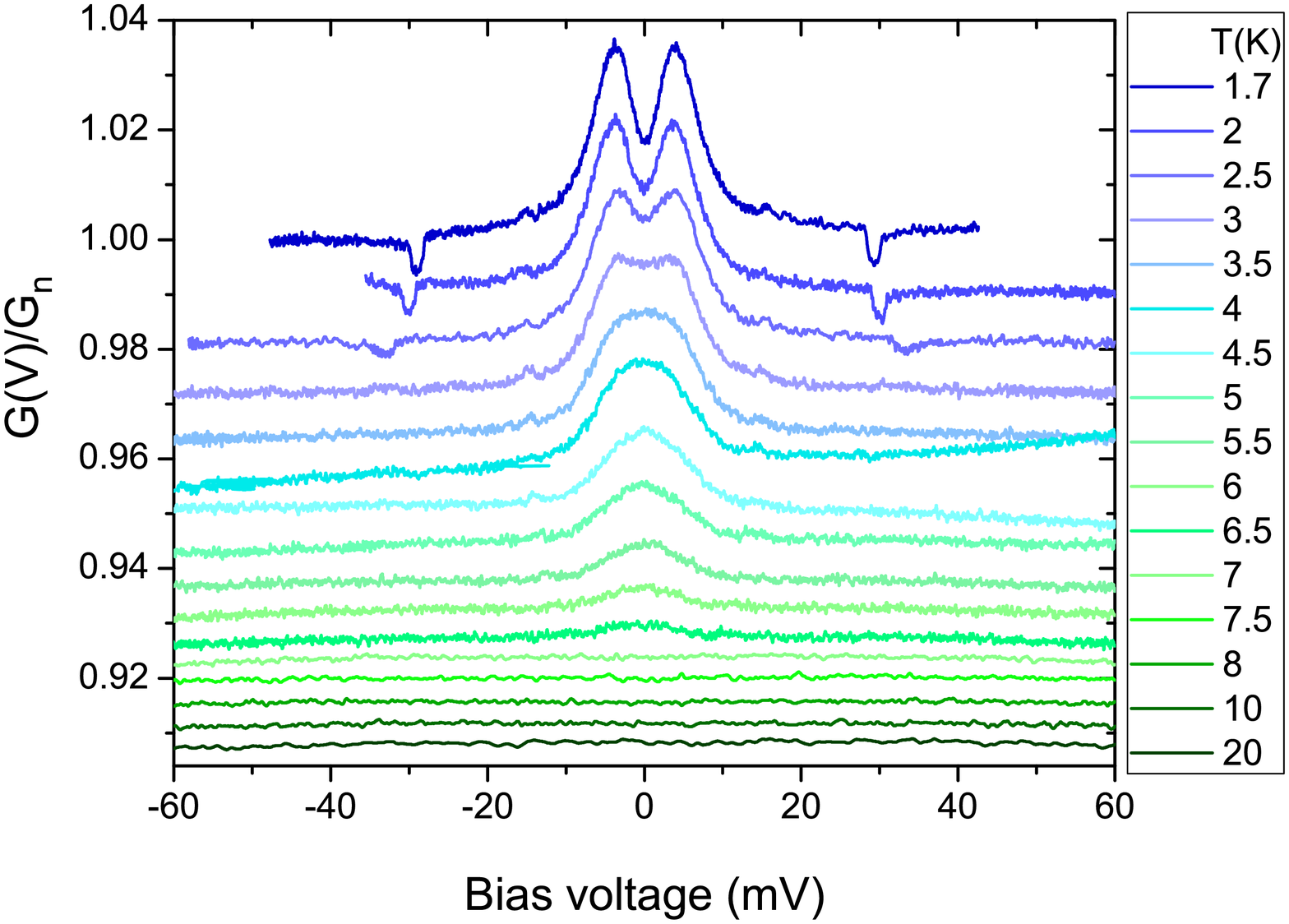}
             \caption{(Color online) Normalized differential conductance measured at different temperatures for a point contact W(Au) - Zr$_{0.97}$V$_{0.03}$B$_{2}$. Curves were vertically shifted by a small amount for  better clarity. In this figure G$_{n}$ represents the differential conductance in the normal state. The symmetrical features at voltages well high of the energy gap are related to resonances or other processes related to the superconducting state \cite{Strijkers,Srikanth, Shan, DeWilde}. Also it is important to mention that the contact differential resistance measured at zero bias voltage was about   50 $\Omega$.}
\label{Fig1.}
\end{figure}

\begin{figure}[h]
	\centering
		\includegraphics[width=0.53\textwidth]{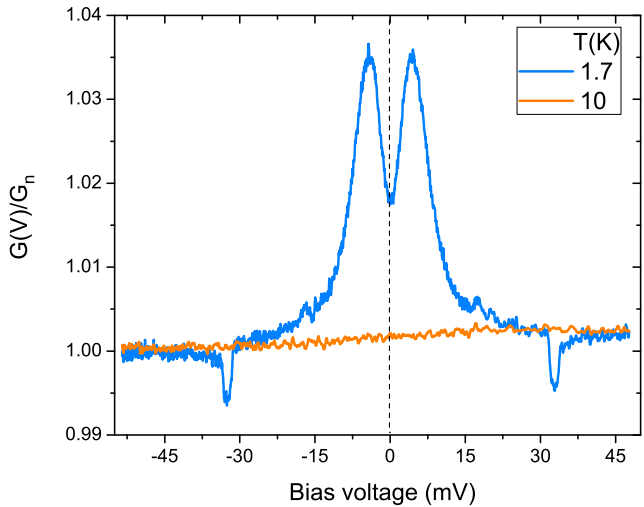}
		   \caption{(Color online) Normalized differential conductance of a junction in  the normal state at  $T$=10 K, and in the superconducting state at $T$=1.7 K. Note the two symmetrical depressions in the superconducting characteristic, these structures and others more, are observed in junctions when are  in the superconducting state, and could be related to resonances of the superconducting characteristic. These features are observed in others studies with PC or tunnel junctions by many workers in the field and are attributed to physical mechanisms not well known in the junctions \cite{Sheet, Strijkers,Srikanth, Shan, DeWilde}.} 
	\label{Fig1.}
\end{figure}

The studies  of the energy gap in Zr$_{0.97}$V$_{0.03}$B$_{2}$  were performed with PCS. The differential conductance, $G(V)= dI/dV$ was obtained by numerical inversion of the differential resistance at different temperatures from 1.7 to 20 K. At low temperature is well known that the differential conductance is directly proportional to the density of electronic states \cite{wolf}. In Fig. 4 we show  results of the measured  characteristics  associated with the energy gap feature. In  this figure close to  6.5 K we see a small hump that rapidly grows as the temperature descends. The hump is dramatically deformed at 3 K  emerging two symmetrically identical peaks around zero bias voltage. These peaks give  a clear evidence of the regime of behavior of the PCS  with this characteristic we calculate the valor of the $Z$ parameter  compared to the shape of the curves with the BTK model. A note of caution must be mentioned; all PCS spectroscopic characteristics reported here were reproducible and with stable characteristics. PCS with values of differential resistance below 20 $\Omega$ were discarded, and only PCS with values above 30 $\Omega$ were used as spectroscopic material. Values of differential resistance above 50 $\Omega$ were not used because presented noisy characteristics. The number of junctions prepared and measured were more than 40 - 47, but the reported data were taken from only the very reproducible junctions. From this we selected the more stables junctions that  show reproducible characteristics.

The size of the energy gap at the minimum accessible temperature of our equipment,  1.7 K, was 2$\Delta \approx$ 7.2 meV obtained by  fitting the experimental data with  BTK model. The shape  of the normalized differential conductance  corresponds to a regime with $Z = 3$ \cite{Strijkers, Lucas, Srikanth}.

In Fig. 5 we  show the normalized differential conductance at two different temperatures,  1.7 and 10 K. Note the dramatic changes when the material is at low temperature in the superconducting state. The feature at 1.7 K is typical of a normal-metal/insulator/superconductor junction in the ballistic regime \cite{Sheet, Gennes, Lucas, Strijkers, Srikanth, Blonder, Jansen, Daghero, Park}.

\subsection{Experimental Data and BTK Model}

 Fig. 6  shows the fitting of experimental data to the BTK model. The  fit and the experimental data are quite similar at low temperatures, from 1.7  to  6 K. At temperatures above 6 K the  feature  associated to the  energy gap  was not fitted, the feature related to  the gap was not seen. The dependence  of gap feature with temperature  was compared with the BCS model, and the fit was performed in ranges from  about 7.5 - 8.5 K, as shown in Fig. 7. It is important to mention that according to our measurements the ratio 2$\Delta$($T$)/K$_{B}$T$_{C}$ extrapolated to zero temperature is approx 8.95, implying a very strong coupling limit for this superconductor \cite{Schmidt}.

\begin{figure}[h]
	\centering
		\includegraphics[width=0.5\textwidth]{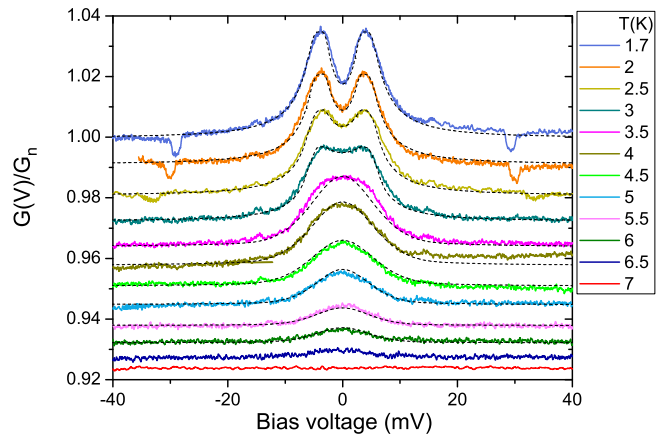}
		   \caption{(Color online) Experimental and theoretical normalized differential conductance as  function of temperature of a point contact at different temperatures. The theoretical fit was performed with the BTK model (dotted lines). The resulting parameters for these data series was $\Delta$ =  3.6 to 3.7, and the $Z$ parameter changes from 2.74 to 3.05.}
	\label{Fig1.}
\end{figure}

\begin{figure}[h]
	\centering
		\includegraphics[width=0.56\textwidth]{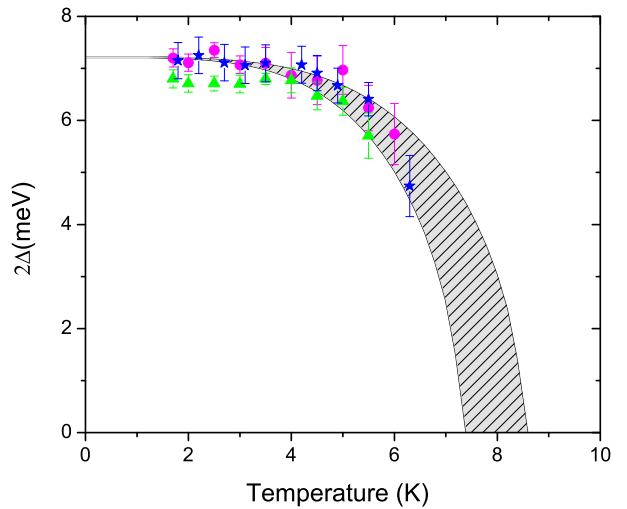}
		   \caption{(Color online) Evolution of the energy gap with temperature, pointed line represents  BCS model, dots represent the variation of the size of the experimental energy gap for some point contacts.}
	\label{Fig7.}
\end{figure}

Despite that unknown size of  the  mean free path of the Zr$_{0.97}$V$_{0.03}$B$_{2}$, we can say that the majority of the point contacts taken into account for this work were operating within the ballistic and diffuse regime. This observation was   based in  comparison between the experimental spectra and the obtained  by the BTK  model. In general, the resistance values of the different point contacts used and measured at zero bias voltage varied from 30 - 50 $\Omega$.

In addition to the determination of the energy gap, we also performed  measurements  at higher bias voltages beyond the feature associated with the superconducting gap,  this is  shown in Fig 4. to 6. Here is shown  features  related to the superconducting state ($\pm$ 33 meV), as possible  resonances of the energy gap. Those characteristics  can not be simulated with the BTK model and  already been observed by other workers in the field \cite{Sheet, Strijkers, Shan, DeWilde, wolf},  the  given explanations are controversial \cite{Shan, Rourke, Sheet, Bashlakov, Hohn, Yansonw}.

\section{Conclusions}

We have presented experimental results of the behavior of the energy gap in the Zr$_{0.97}$V$_{0.03}$B$_{2}$ superconducting compound. Analysis of the experimental data using the BTK theoretical model give us values of the size of the energy gap which at low temperature was determined to be  2$\Delta = 7.2 $ meV, also we determined the  value for the ratio 2$\Delta$(0)/K$_{B}$T$_{C}$ $\approx$ 8.95 which suggests a  superconductor in the strong coupling regime.

\begin{acknowledgments}

This work was partially supported by grants by DGAPA-UNAM project IN106014, CONACyT M\'{e}xico.  L\'{o}pez-Romero thanks to CONACyT-M\'{e}xico for a scholarship and DGAPA-UNAM, IN106014. We also thanks  Joaqu\'{i}n Morales,  Alberto L\'{o}pez, R. Reyes,  and F. Silvar for technical support and helium provisions.

\end{acknowledgments}

\end{document}